\documentclass[pra,twocolumn,groupedaddress,shownopacs,amssymb]{revtex4-1}
\usepackage{graphicx,psfrag,amsmath,amssymb}
\usepackage[usenames, dvipsnames]{color}
\usepackage{braket}
\usepackage{amsmath}
\usepackage{amssymb}
\usepackage{gensymb}
\usepackage{makeidx}
\usepackage{mathtools}
 
\usepackage[normalem]{ulem}
\usepackage{float}
\usepackage[T1]{fontenc} 
\usepackage{balance}
\usepackage{flushend}
\usepackage{lineno}
\usepackage{epsfig}
\usepackage{subfigure}
\usepackage{mathtools}
\usepackage{setspace}
\usepackage{bm}
\usepackage{ulem}
\usepackage{times}
\usepackage{enumitem}
\usepackage{xcolor}
\usepackage{multirow}
\usepackage{soul}
\usepackage{dcolumn}
\usepackage{balance}
\usepackage{epstopdf}
\usepackage{braket}
\usepackage{color}
\usepackage{diagbox}
\usepackage[utf8]{inputenc}
\usepackage{hyperref}
\hypersetup{
    colorlinks=true,
    linkcolor=Blue,
    citecolor=Blue,
    filecolor=Blue,
    urlcolor=Blue,
    }
\usepackage{lineno}
\begin{document}
\title{Quasiparticle spectra of mixtures of dipolar and non-dipolar condensates at zero and finite temperatures}
\author{Harsimranjit Kaur}
\affiliation{Department of Physics, Central University of Rajasthan, Ajmer - 305817, India}
\author{Kuldeep Suthar}
\email[Corresponding author. ]{kuldeep.suthar@curaj.ac.in}
\affiliation{Department of Physics, Central University of Rajasthan, Ajmer - 305817, India}

\date{\today}
%%%%%%%%%%%%%%%%%%%%%%%%%%%%%%%%%%%%%%%%%%%%%%%%%%%%%%%%%%%%%%%%%%%%%%%%%%%%%
%%%%%%%%%%%%%                  Abstract                        %%%%%%%%%%%%%%
%%%%%%%%%%%%%%%%%%%%%%%%%%%%%%%%%%%%%%%%%%%%%%%%%%%%%%%%%%%%%%%%%%%%%%%%%%%%%
\begin{abstract}
	We examine the low-lying collective quasiparticle modes of a mixture of flattened infinite-pancake Bose-Einstein condensates having dipolar and non-dipolar atomic species. The dipolar atomic species have permanent magnetic dipolar moments. We employ Hartree-Fock-Bogoliubov theory to investigate the distinct axial collective spectra at zero and finite temperatures corresponding to phase separation phenomena stemming from the dipole-dipole interaction of dipolar atomic species. When the dipolar interaction is tuned to be repulsive, the number of zero-energy axial modes decreases, reflecting the system's tendency towards mixing. For a large number of atoms per unit area, we show that the attractive (repulsive) dipolar interaction strengths lead to ground states with non-dipolar (dipolar) atomic species at the periphery, and this leads to a discontinuity in quasiparticle mode evolution. We finally reveal that miscibility driven by thermal fluctuations at finite temperatures exhibits dipole mode hardening of axial excitations, confirmed by the loss of long-range phase coherence through the correlation function. The mode mixing in the dispersion relations ascertains a dipolar strength-dependent miscibility transition and the low-lying quasiparticle mode evolution. 

\end{abstract}
\maketitle

%%%%%%%%%%%%%%%%%%%%%%%%%%%%%%%%%%%%%%%%%%%%%%%%%%%%%%%%%%%%%%%%%%%%%%%%%%%%%
%%%%%%                   Section: Introduction                         %%%%%%
%%%%%%%%%%%%%%%%%%%%%%%%%%%%%%%%%%%%%%%%%%%%%%%%%%%%%%%%%%%%%%%%%%%%%%%%%%%%%

\section{Introduction}\label{Introduction}
 The interatomic interactions play a vital role in determining the properties of the quantum degenerate atomic gases. While most of the ultracold experiments have been dominated by short-range interactions, over the decades, significant progress in the realization of Bose-Einstein condensates (BECs) of atoms with large magnetic dipole moments, including $^{52}$Cr~\cite{griesmaier_06,koch_08,beaufils_08,depaz_13}, $^{164}$Dy~\cite{lu_11,preti_24}, and $^{168}$Er\cite{aikawa_12,ulitzsch_17}, as well as the realizations of heteronuclear molecules~\cite{ni_08,bigagli_24}, have sparked growing interest in the physics of dipole-dipole interactions (DDIs). The dipolar interaction is long-range and anisotropic, leading to novel spatially modulated quantum states such as supersolid~\cite{boninsegni_12,tanzi_19,natale_19,chomaz_19,bottcher_21,hertkorn_21,bland_22,norcia_22,elena_23}, self-bound
droplet~\cite{bulgac_02,chomaz_16,schmitt_16,barbut_16,bottcher_19,schmidt_22}, and pattern formation in ferrofluid~\cite{lahaye_09,kadau_16,hertkorn_21a}. These phases surpass the dipolar instability and are stabilized by the inevitable role of quantum fluctuations. The dipolar interaction influences the macroscopic transport properties~\cite{wenzel_18,he_25}, collective excitation spectrum~\cite{ralf_06,bijnen_10,bismut_10,lima_12}, and exhibits a roton-maxon excitation spectrum~\cite{santos_03,fischer_06,bisset_13,chomaz_18}. The anisotropy of the interaction results in the distinct condensate shapes and properties that depend on the trap geometry and dipole orientations~\cite{xi_11,gligoric_10}. In lower-dimensional traps, the tuning of the orientation of the dipoles via external fields becomes especially critical in determining the behavior of the condensates~\cite{lu_10} and results in quantum phase transitions~\cite{lahaye_09}.\\

The introduction of a second species of atoms, two-component Bose-Einstein condensates (TBECs), provides a rich platform for exploring diverse phenomena in physics. These include suppression of collapse by quantum fluctuations and demixing effects in short-range binary mixtures~\cite{petrov_15,petrov_16}. The miscibility and phase separation phenomena of mixtures with contact interaction have been explored in detail both theoretically and experimentally. The recent experimental realization of dipolar mixtures~\cite{trautmann_18,durastante_20,politi_22} paves a way to explore the intriguing effects of long-range and anisotropy on the mixing properties of TBECs. The dipole orientations create a versatile system to feature remarkable properties, and provide a new perspective for atomtronic applications~\cite{hector_25,luigi_22,amico_21}. The different mutual orientations of the dipole moments influence the miscibility of the binary mixtures. In particular, for equal magnitudes of dipolar interactions, the opposite direction of the polarization induces miscibility transitions in contrast to the same orientations~\cite{gligoric_10,lee_21,madronero_25}. During the miscibility transition, the dipolar mixtures exhibit transient structure formation~\cite{young_12,kumar_17}. The strength of the dipolar interaction controls the type of phase separation~\cite{wang_18,tomio_20}. Of particular interest, the mixtures of dipolar and non-dipolar species exhibit interfacial pattern formation~\cite{saito_09} and stabilize the formation of the supersolid phase even in the absence of beyond-mean-field correction~\cite{li_22}. In the phase-separated state, the attractive (repulsive) dipolar interaction strength pushes out non-magnetic (magnetic) atoms of the species~\cite{xi_11}. Hence, the interplay of contact and nonlocal dipolar interactions determines the spatial density distribution of the ground states in the binary mixtures. The study of collective quasiparticle excitations is an excellent tool to analyze the interplay of short- and long-range interactions in many-body systems. The excitations of trapped BECs with short-range interaction are well understood~\cite{mukherjee_25}; however, the excitations of mixtures of dipolar and non-dipolar atomic condensates are relatively less explored.  

We investigate the evolution of quasiparticles as a quantum mixture is driven towards the (im)miscibility by dipolar interaction and further explore thermal effects, revealing intriguing phase transitions driven by temperature in infinite pancake Bose-Einstein condensates. Owing to translational invariance in the radial directions, the quasiparticle modes with different transverse momentum get decouple. Here, we examine the axial quasiparticle excitations corresponding to zero in-plane momentum. We first examine the quasiparticle mode hardening and discontinuity in the excitation spectrum with dipolar interaction strength at zero temperature. The finite-temperature study reveals the robustness of the mixing transition against thermal fluctuations. We further examine the loss of phase coherence through the correlation function. Finally, the miscibility and mode mixing due to dipolar interaction are confirmed by the dispersion relations of the binary mixtures.       

The paper is organized as follows: In section~\ref{quasi-1d}, we present the Hartree-Fock-Bogoliubov formalism applied to a binary mixture of dipolar and non-dipolar condensates. Section~\ref{RESULTS AND DISCUSSIONS} discusses the evolution of quasiparticle modes by varying the ratio of long- and short-range interactions at zero and finite temperatures. We further discuss the correlation functions and dispersion relation. Finally, section~\ref{CONCLUSIONS} provides a summary of the key findings and conclusions.

%%%%%%%%%%%%%%%%%%%%%%%%%%%%%%%%%%%%%%%%%%%%%%%%%%%%%%%%%%%%%%%%%%%%%%%%%%%%THEORY%%%%%%%%%%%%%%%%%%%%%%%%%%%%%%%%%%%%%%%%%%%%%%%%%%%%%%%%%%%%%%%%%%%%%%%%%%%%%%%%%%%%%%%%%%%%%%%%%%%%%%%%
\section{Infinite pancake regime and HFB approximation}
\label{quasi-1d}
 We consider a binary mixture consisting of dipolar and non-dipolar Bose-Einstein condensates, where the long-range dipole-dipole interaction between two particles is given by $\Phi_{dd}(\mathbf{r}) = \int d^3 \mathbf{r'} \, U_{dd}(\mathbf{r} - \mathbf{r}') |{\psi}(\mathbf{r}')|^{2}$. $U_{dd}(\mathbf{r}) = (C_{dd}/4\pi)~[\hat{e}_i \hat{e}_j (\delta_{ij} - 3\hat{r}_i \hat{r}_j)/r^3]$ is the interaction between two dipoles that are polarized by an external field along a unit vector $\hat{\mathbf{e}}$ and separated by a distance $\mathbf{r}$. \( C_{dd} \) is the strength of the dipolar interaction~\cite{kanjilal_07, lahaye_09}. Here, $|{\psi}(\mathbf{r})|^{2} = n_{c}(\mathbf{r})$ is the density of condensate atoms. The short-range intracomponent contact interaction is $U_k = {4\pi \hbar^2 a_k}/{m_k}$ with $k = 1,2$ as the species index and $a_k$ is the $s$-wave scattering length of the $k$th species. The intercomponent interaction is $U_{12}={2\pi \hbar^2 a_{12}}/{m_r}$, where $m_r =m_1 m_2 / (m_1 + m_2)$ is the reduced mass and $a_{12}$ is the intercomponent scattering length. The non-dipolar component interacts solely via short-range contact interactions. The interplay between contact and dipolar interactions (of the dipolar component) governs the collective behavior and stability of the mixture. We assume the dipolar component as the first species while the non-dipolar component as the second species. The strength of the dipolar interaction relative to the contact interaction is characterized by a dimensionless parameter \( \varepsilon_{dd} \equiv C_{dd}/3U_{1} \)~\cite{dell_04,eberlein_05}. We consider repulsive contact interaction $U_{k}>0$ such that the negative $\varepsilon_{dd}$ corresponds to $C_{dd}<0$. The amplitude and sign of $C_{dd}$ can be tuned by rotation of the polarization axis~\cite{giovanazzi_02,madronero_25}. For dipoles aligned along the $z$-direction, the dipolar potential can be expressed in terms of a dimensionless parameter as $\Phi_{dd}(\mathbf{r}) = -U_{1}\varepsilon_{dd}[3\partial^{2}_{z}\phi(\mathbf{r}) + n_c(\mathbf{r})]$ with $\phi(\mathbf{r})$ being the analogous electrostatic potential satisfying the Poisson equation $\nabla^{2}\phi = - n_c(\mathbf{r})$. Here, the first term of $\Phi_{dd}$ is anisotropic and long range, while the second term is short range and contact-like. 
 
 We consider a condensate mixture by neglecting the radial trapping frequencies and enhancing the axial confinement, thereby creating a highly flattened, infinite pancake-shaped geometry with uniform radial density~\cite{parker_09,xi_11}. In this limit, the Poisson equation reduces to ${\partial^{2} \phi}/{\partial z^{2}} = -\, n_{c}(z),$ and the dipolar interaction simplifies to a contact-like form, $\Phi_{dd}(z) = 2 U_{1}\, \varepsilon_{dd}\, |\psi_{1}(z)|^{2}.$ In this particular geometry, the uniform radial density results in vanishing non-local radial dipolar terms. In such highly flattened pancake condensates, the dipolar contribution to the mean-field interaction reduces entirely to a local term~\cite{parker_08,lee_22}. The Heisenberg equation of motion for the Bose field operator is given by
\begin{subequations}
\begin{eqnarray}
i \hbar \frac{\partial \hat{\Psi}_1}{\partial t} &=& \left(-\frac{\hbar^2}{2m_1} \frac{\partial^2}{\partial z^2} + V_{1} + U^{\rm eff}_1 \hat{\Psi}_1^\dagger \hat{\Psi}_1 
 + U_{12} \hat{\Psi}_2^\dagger \hat{\Psi}_2 \right) \hat{\Psi}_1,~~~~~~~~\\
i \hbar \frac{\partial \hat{\Psi}_2}{\partial t} &=& \left(-\frac{\hbar^2}{2m_2} \frac{\partial^2}{\partial z^2} + V_{2} + U_2 \hat{\Psi}_2^\dagger \hat{\Psi}_2 
+ U_{12} \hat{\Psi}_1^\dagger \hat{\Psi}_1 \right) \hat{\Psi}_2,~~~~~~~~
\end{eqnarray}
\end{subequations}
where $V_{k}(z) = m_{k} \omega_{kz}^{2} z^{2}/2$ is an external harmonic trapping potential and $U^{\rm eff}_{1} = (1 + 2\varepsilon_{dd})U_1$. The effective interaction modifies the critical intercomponent scattering length of phase separation~\cite{xi_11}. It is worth noting that the $\varepsilon_{dd}$ varied from $-0.5$ to $0.5$ in the present work. The minimum value $\varepsilon_{dd} = -0.5$ leads to zero effective interaction $U^{\rm eff}_{1}$. Therefore, the system remains stable and does not collapse for the choice of $\varepsilon_{dd}$, and a
beyond mean-field Lee–Huang–Yang (LHY) correction~\cite{petrov_15,petrov_16} is not included.  Under the Hartree-Fock-Bogoliubov (HFB) approximation, the field operator can be expressed as the sum of the mean-field and the fluctuations over it, $\hat{\Psi}_k (z,t) = \psi_k (z) + \hat{\varphi}_k (z,t)$ with $k=1,2$ as the species index. Here, $\psi_k$ represents the ground state condensate wave-function and $\hat{\varphi}_{k}$ is the fluctuation operator that incorporates quantum and thermal fluctuations at zero and finite temperatures, respectively. The equation of motion for the condensates is given by the coupled generalized Gross-Pitaevskii equations (GPEs), given as
\begin{subequations}
  \begin{eqnarray}
    \left( \frac{-\hbar^2}{2m_1} \frac{\partial^2}{\partial z^2} + V_{1} - \mu_1 \right) \psi_1 &+& U^{\rm eff}_1 \left( n_{c1} + 2 \tilde{n}_1 \right) \psi_1 
    + U_{12} n_2 \psi_1 = 0, \nonumber\\~\\
    \left( \frac{-\hbar^2}{2m_2} \frac{\partial^2}{\partial z^2} + V_{2} - \mu_2 \right) \psi_2 
&+& U_{2} \left( n_{c2} + 2 \tilde{n}_2 \right) \psi_2 + U_{12} n_1 \psi_2 = 0, \nonumber\\
\end{eqnarray}
\label{coupled_gpes}
\end{subequations}
where $\tilde{n}_{k} = \langle\hat{\varphi}^{\dagger}_{k}\hat{\varphi}_{k}\rangle$ is the density of the excited non-condensate population and $n_{k} = |\psi_{k}|^{2} + \tilde{n}_{k}$ is the total density of the $k$th component. The equation of motion of the fluctuation operator for the first and second species is
\begin{subequations}
\begin{eqnarray}
  i\hbar\frac{\partial\hat{\varphi}_1}{\partial t} &=& 
   \bigg[-\frac{\hbar^2}{2m_1} \frac{\partial^2}{\partial z^2} + V_{1} + 2 U^{\rm eff}_1 (n_{c1} + \tilde{n}_1) 
- \mu_1 + U_{12} {n}_2 \bigg]\hat{\varphi}_1 \nonumber \\ 
 &+& U^{\rm eff}_1 \psi^{2}_1 \hat{\varphi}_1^\dagger + U_{12} \psi_2^* \psi_1 \hat{\varphi}_2 
+ U_{12} \psi_1 \psi_2 \hat{\varphi}_2^\dagger, \\
    i \hbar \frac{\partial\hat{\varphi}_2}{\partial t} &=& 
\bigg[-\frac{\hbar^2}{2m_2} \frac{\partial^2}{\partial z^2} + V_{2} + 2U_{2} (n_{c2} + \tilde{n}_2) - \mu_2 
+ U_{12} {n}_1 \bigg]\hat{\varphi}_2 \nonumber \\  
&+& U_{2} \psi_2^2\hat{\varphi}_2^\dagger + U_{12} \psi^{*}_1 \psi_2 \hat{\varphi}_1 + U_{12} \psi_1 \psi_2 \hat{\varphi}_1^\dagger,
\end{eqnarray}
\label{fluc}
\end{subequations}
where the three-body correlation terms and the anomalous density term $\xi_{k} = \psi^{2}_{k} + \tilde{\xi}_{k}$ with $\tilde{\xi}_{k} = \langle\hat{\varphi}_{k}\hat{\varphi}_{k}\rangle$ are omitted. The latter leads to a finite energy gap in the excitation spectrum and thus results in a gapped spectrum~\cite{griffin_96}. The self-consistent calculation of $\tilde{n}$ leads to a gapless excitation spectrum of the binary atomic mixtures.

Under Bogoliubov transformation, the fluctuation operators can be written in terms of quasiparticle mode functions as
\begin{equation}
  {\hat{\varphi}}_k = \sum_j \left( u_{kj} \hat{\alpha}_j e^{-i E_j t / \hbar} - v_{kj}^* \hat{\alpha}_j^\dagger e^{i E_j t / \hbar} \right),
  \label{bogo_trans}
\end{equation}
where $u_{kj}$ and $v_{kj}$ are the quasiparticle amplitudes corresponding to the $k$th species, $E_{j}$ is the quasiparticle energy of the $j$th mode, and $\hat{\alpha}_{j}(\hat{\alpha}_j^\dagger)$ is the quasiparticle annihilation (creation) operator that satisfies the Bose commutation relations. 
It is worth noting that for zero in-plane momentum, the dipolar interaction becomes momentum independent and can be approximated to local contact-like interactions. Here, we consider the axial excitations with mode energies and amplitudes being functions of $z$. Using the above transformation [Eq.~(\ref{bogo_trans})] in Eq.~(\ref{fluc}), we obtain the coupled HFB equations. The coupled eigenvalue equations can be written in the matrix form as
\begin{equation}
\begin{pmatrix}
\mathcal{A}_1 & \mathcal{B}_1 & \mathcal{C}_1 & \mathcal{C}_2 \\
\mathcal{- B}_1^{*} & \mathcal{-A}_1^{*} & \mathcal{-C}_2^{*} & -\mathcal{C}_1^{*} \\
\mathcal{C}_1^{*} & \mathcal{C}_2 & \mathcal{A}_2 & \mathcal{B}_2 \\
\mathcal{- C}_2^{*} & \mathcal{-C}_1 & \mathcal{-B}_2^{*} & \mathcal{-A}_2{*}
\end{pmatrix}
\begin{pmatrix}
u_{1j}\\
v_{1j}\\
u_{2j}\\
v_{2j}
\end{pmatrix}=\omega
\begin{pmatrix}
u_{1j}\\
v_{1j}\\
u_{2j}\\
v_{2j}
\end{pmatrix},
\label{bdg_matrix}
\end{equation}
where the matrix elements are given by
\begin{subequations}
\begin{align*}
\mathcal{A}_1 &= \hat{h}_1 + 2U^{\rm eff}_{1} n_1 + U_{12} n_2,
\qquad
\mathcal{B}_1 = - U^{\rm eff}_1 \psi_1^2, \\
\mathcal{C}_1 &= U_{12} \psi_1 \psi_2^*, \qquad
\mathcal{C}_2 = - U_{12} \psi_1 \psi_2, \\
\mathcal{A}_2 &= \hat{h}_2 + 2U_2 n_2 + U_{12} n_1, 
\qquad
\mathcal{B}_2 = - U_2 \psi_2^2.
\end{align*}
\label{bdg2s}
\end{subequations}
Here, $\hat{h}_{k} = \left(-\hbar^2/2m_{k}\right)\partial^2/\partial z^2 + V_{k}(z) - \mu_{k}$. The above HFB equations are solved self-consistently to obtain the quasiparticle mode functions and corresponding mode energies. The noncondensate components, or the total sum of the thermal and quantum fluctuations for each species, are 
\begin{equation}
  \tilde{n}_k = \sum_j \left[ \left( |u_{kj}|^2 + |v_{kj}|^2 \right) \langle\alpha^{\dagger}_{j}\alpha_{j}\rangle + |v_{kj}|^2 \right],
  \label{non-cond}
\end{equation}
where $\langle\alpha^{\dagger}_{j}\alpha_{j}\rangle \equiv N_{0}(E_{j}) = (e^{E_{j}/k_{B}T} - 1)^{-1}$ is the Bose factor of the quasiparticle state with excitation energy $E_{j}$ at temperature $T$. The coupled equations [Eqs.~(\ref{coupled_gpes}) and Eqs.~(\ref{bdg2s})] are self-consistently solved until the solutions converge to the required level of accuracy. 
%%%%%%%%%%%%%%%%%%%%%%%%%%%%%%%%%%%%%%%%%%%%%
%%%%%        Results and Discussions    %%%%%
%%%%%%%%%%%%%%%%%%%%%%%%%%%%%%%%%%%%%%%%%%%%%
\section{RESULTS AND DISCUSSIONS}\label{RESULTS AND DISCUSSIONS}
We first examine the evolution of low-lying quasiparticle modes as a function of the dipolar interaction parameter. The effects of quantum fluctuations present at zero temperature are incorporated in computing the collective excitation energies. The radial translational invariance allows decoupling of modes with different transverse momentum~\cite{fischer_06,lima_12}. In the following, we restrict the investigation to zero in-plane momentum corresponding to purely axial collective modes. The restrictions to axial mode sectors constitute a mode selection based on momentum-sector decoupling in the translationally invariant system.
We consider a binary mixture consisting of \(^{52}\mathrm{Cr}\) dipolar atoms as the first species, with a scattering length of \(5\,\mathrm{nm}\)~\cite{griesmaier_06}, and non-dipolar \(^{87}\mathrm{Rb}\) atoms as the second species, with a scattering length of \(10\,\mathrm{nm}\)~\cite{theis_04}, confined in an infinitely extended radial trap. We choose the trapping frequency such that $m_{\rm rat} \omega_{\rm rat}^{2} = 1$, where $m_{\rm rat}=m_2/m_1$ and $\omega_{\rm rat}=\omega_{2z}/\omega_{1z}$ fixing the harmonic confinement in dimensionless units. The trapping frequency of the first component is set to $\omega_{1z} = 2\pi \times 160\,\mathrm{Hz}$. The harmonic oscillator length $a_{\mathrm{osc}} = \sqrt{\hbar/(m_{1}\omega_{1z})}$ and energy quanta $\hbar \omega_{1z}$ correspond to the length and energy scale of the system. The ground-state wave function is obtained by solving the coupled GPEs using the split-step Crank-Nicolson method with imaginary-time propagation. In the numerical computations, we employ a spatial step size $\Delta z = 0.025$, a temporal step size $\Delta t = 0.0001$, and the number of grid points as $2500$. The number of grid points and spatial spacing are chosen such that the condensate lies well within the grid. The convergence of the solution is ensured by the values of ground-state energy that converge up to a tolerance of $10^{-6}$. To determine the excitation spectrum, we cast the coupled HFB equations [Eq.~(\ref{bdg2s})] as a matrix eigenvalue problem in the $200$ harmonic oscillator basis and diagonalize the matrix using the LAPACK routine {\tt zgeev}. The eigenvalues of the problem are quasiparticle energies $E_{j}$'s and eigenfunctions are the mode amplitudes, $u_{k}$'s and $v_{k}$'s. For finite temperature calculations, the non-condensate densities [Eq.~(\ref{non-cond})], ground-state wave-functions, and chemical potentials are updated, and iterations are continued until desired convergence in the number of condensate and non-condensate atoms is attained. In the presence of the fluctuations, the convergence of the self-consistent solution sometime becomes unstable, with frequent and pronounced oscillations. To mitigate this issue and enhance convergence stability, we apply a \textit{successive under-relaxation} technique, expressed as ${\tilde{n}}_{\rm IC}^{\text{new}}(z) = \gamma_{{\rm un}}{\tilde{n}}_{\rm IC}(z) + \left(1 - {\gamma}_{\rm un}\right){\tilde{n}}_{{\rm IC}-1}(z)$, where \(\gamma_{{\rm un}}\) is the under-relaxation parameter and $\rm IC$ stands for iteration cycle. Here, we set $\gamma_{\rm un} = 0.5$.

%%%%%%%%%%%%%%%%%%%%%%%%%%%%%%%%%%%%%%%%%%%%%%%%%%%
%%%%%            Mode evolution at T = 0 K %%%%%%%%
%%%%%%%%%%%%%%%%%%%%%%%%%%%%%%%%%%%%%%%%%%%%%%%%%%%
\subsection{Quasiparticle mode evolution at $T = 0$K}
\subsubsection{Dipolar-interaction-driven miscibility transition}
We discuss the effects of dipolar strength on the low-lying excitations of a binary mixture of dipolar and non-dipolar condensates. The system exhibits the phase separation phenomenon when $U_{12}$ exceeds the geometric mean of $U^{\rm eff}_{1}$ and $U_{2}$. We consider the number of atoms per unit area for each species is $100$. We begin with a phase-separated configuration, for which the intercomponent scattering length is set to \(a_{12} = 7\,\mathrm{nm}\). The binary condensates exhibit two gapless zero-energy Goldstone modes arising from two spontaneous gauge symmetry breakings.

It is important to note that the system also possesses gapless in-plane excitations, including roton modes due to finite in-plane momenta~\cite{santos_03,fischer_06}. The number of Goldstone modes further depends on the intercomponent interaction and consequently on the ground state density distribution profiles. 
Since the $s$-wave scattering length controls the dimensionless dipolar strength and the anisotropic nature of the dipolar interaction allows 
$\varepsilon_{dd}$ to be of negative sign. 
\begin{figure}[h]
    \includegraphics[width=\linewidth]{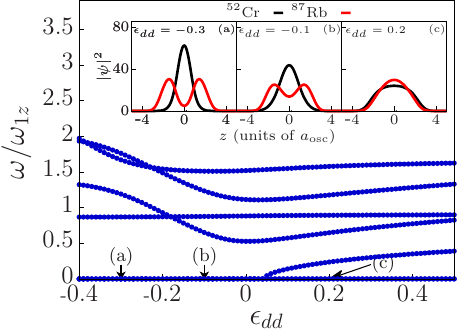}
    \caption{The evolution of low-lying quasiparticle modes energies as a function of the dipole-dipole interaction strength parameter $\varepsilon_{dd}$ for 
the intercomponent scattering length  $a_{12} = 7$ nm. The evolution is shown for zero temperature by incorporating the quantum fluctuations. The inset panels display the corresponding density profiles for three representative values, $\varepsilon_{dd} = -0.3, -0.1,$ and $0.2$, illustrating the transition from immiscible to miscible phases. These specific values are also indicated by arrows in the main figure. The quasiparticle energies are scaled in units of the harmonic oscillator energy.}
    \label{mode_a12_7nm}
\end{figure}
When $\varepsilon_{dd} = -0.4$, the ground state of the system acquires a sandwich profile, in which, due to attractive interaction, Cr remains at the center and is flanked by Rb atoms. As $\varepsilon_{dd}$ varies from negative to positive, the effective interaction of Cr becomes repulsive, and the system gets mixed. Thus, the variation of dipolar interaction $\varepsilon_{dd}$ leads to a quantum phase transition from an immiscible to a mixed state, which is evident from the inset of Fig.~\ref{mode_a12_7nm}, where the density profiles are shown for $\varepsilon_{dd} = -0.3, -0.1,$ and $0.2$.

For attractive dipolar interaction and \(\varepsilon_{dd} \lesssim 0.06\), the system resides in the immiscible phase. In this regime, the ground state is equivalent to three distinct condensate fragments, and thus the excitation spectrum possesses three zero-energy modes. In the sandwich configuration, phase-separation causes additional spatial symmetry breaking as a result of the formation of interfaces. There are two interfaces on either side of the inner component. The position of the interfaces can be moved without energy cost, and thus the interfacial (breathing) motion leads to an additional gapless excitation. A small shift in the interface does not change the interaction and trap energy to the leading order; however, it modifies the density profiles. It follows from the Goldstone theorem that state breaking of symmetry of the ground state of the Hamiltonian leads to gapless excitations or zero-energy Goldstone 
modes~\cite{goldstone_61,goldstone_62,takeuchi_13,takahashi_15,roy_14,suthar_15}. The lowest finite-energy mode is the Kohn mode, which remains steady in energy with change in $\varepsilon_{dd}$. As \(\varepsilon_{dd}\) exceeds $0.04$, the third (additional) zero-energy mode gains energy. It is due to the fact that the system approaches miscibility where both species overlap and correspond to two zero-energy modes. Hence, the immiscible-miscible phase transition is reflected in the excitation spectra as a hardening of one of the zero-energy modes. The structural transformation of the low-lying modes validates the change in quasiparticle energy with $\varepsilon_{dd}$. 
\begin{figure}[h]
    \includegraphics[width=\linewidth]{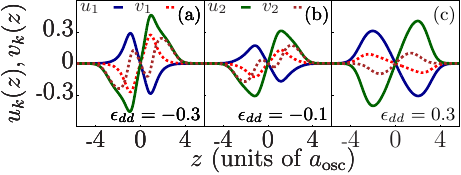}
    \caption{The evolution of the quasiparticle amplitudes associated with the zero-energy mode for a binary BECs including quantum fluctuations, 
	     with a fixed $a_{12} = 7$ nm. As the dipole-dipole interaction shifts from attractive to repulsive, the zero-energy mode of 
	     $^{87}\text{Rb}$ transforms to a dipole mode in the miscible phase.}
    \label{mode_fn_a12_7nm}
\end{figure}
At attractive $\varepsilon_{dd}$, the softened out-of-phase dipole mode corresponding to the sandwich density profile possesses zero energy. This mode remains structurally similar and gains in energy and transforms into an out-of-phase dipole of mixed overlapped density configuration, as the evolution is depicted in Fig.~\ref{mode_fn_a12_7nm}(a-c). 
%%%%%%%%%%%%%%%%%%%%%%%%%%%%%%%%%%%%%%%%%%%%%%%%
%%% Position swapping (exchange) of species %%%%
%%%%%%%%%%%%%%%%%%%%%%%%%%%%%%%%%%%%%%%%%%%%%%%%
\subsubsection{Position swapping of species}
 We further discuss a remarkable phenomenon of phase-swapping, where the positions of the dipolar and non-dipolar components exchange positions with respect to the trap center. In particular, when the dipolar strength is small (large), the non-dipolar (dipolar) component of the mixture is pushed out~\cite{xi_11}. The swapping occurs for a larger number of atoms per unit area or, consequently, for strong repulsive contact interactions. Here, we consider \(1000\) atoms per unit area in each component and an intercomponent scattering length as \(a_{12} = 12\,\text{nm}\). Due to strong intercomponent interaction, the system remains in the immiscible regime, where one species is flanked by the other. We examine the role of $\varepsilon_{dd}$ in swapping phenomena on low-energy collective excitations.
\begin{figure}[h]
    \includegraphics[width=\linewidth]{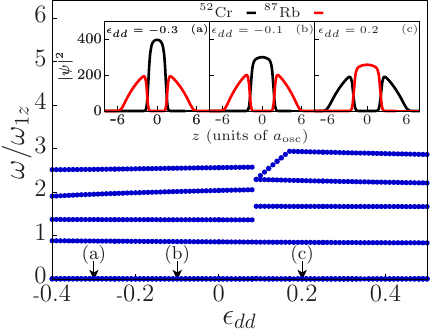}
    \caption{Low-lying quasiparticle mode energies obtained using the HFB-Popov approximation, as a function of the dipole-dipole interaction strength parameter $\varepsilon_{dd}$ for $a_{12} = 12$ nm. 
	The inset plots show the density profiles at three representative values, $\varepsilon_{dd} = -0.3, -0.1,$ and $0.2$, illustrating the phase swapping of the species while remaining in the immiscible regime.}
    \label{mode_a12_12nm}
\end{figure}
For negative values of \(\varepsilon_{dd}\), corresponding to attractive dipole-dipole interactions, the dipolar species minimizes its energy by localizing in the center of the harmonic trap, where the density is maximum. As \(\varepsilon_{dd}\) increases and crosses into the repulsive regime (\(\varepsilon_{dd} > 0\)), the nature of the dipole-dipole interaction changes dramatically. Once the repulsion becomes sufficiently strong, typically for \(\varepsilon_{dd} \gtrsim 0.1\), the dipolar component experiences a net outward force. This leads to a spatial reconfiguration: the non-dipolar species now occupies the central region of the trap, while the dipolar species is expelled to form a shell around it. This is evident from the inset density profiles in Fig.~\ref{mode_a12_12nm}, shown for $\varepsilon_{dd} = -0.3, -0.1,$ and $0.2$.

The energy of the first excited Kohn mode associated with center-of-mass oscillations remains essentially invariant across the transition. This is consistent with the generalized Kohn theorem~\cite{kohn_61}, which ensures that, in harmonic traps, center-of-mass modes decouple from interparticle interactions. We verified that the Kohn mode function structurally remains similar with the change in $\varepsilon_{dd}$. In contrast, the quadrupole mode, which involves deformations of the density profile shape rather than simple translations, exhibits a clear discontinuity in its excitation energy near the critical value of \(\varepsilon_{dd}\). This phase transition reflects the system's sensitivity to internal structural changes. The shift in quadrupole mode energy corresponds directly to the spatial inversion of the two components, signaling a first-order-like transition between two distinct immiscible states with reversed core-shell geometries. 

%%%%%%%%%%%subsection:Mode evolution of the trapped binary BECs at finite Temperature%%%%%%%%%%%%%%%%%%%%%%%%%%%%%%%%%%%%%%%%%%%%%%%%%%%%%%%%%%%%%%%%%%%%%%%%%%%%%%%%%%%
\subsection{Mode evolution of the trapped binary BECs at finite temperatures}
\begin{figure}[h]
    \includegraphics[width=\linewidth]{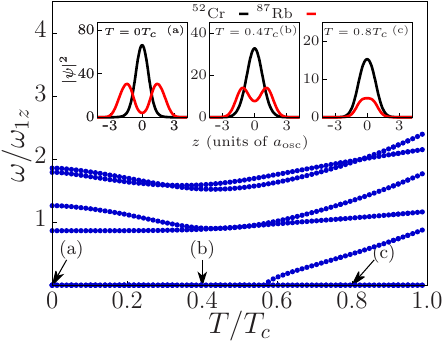}
	\caption{Evolution of the low-lying quasiparticle mode energies as a function of temperature for a fixed attractive dipolar interaction strength, $\varepsilon_{dd} = -0.35$, with $N_1 = N_2 = 100$ (number of atoms per unit area). As the temperature increases, the non-condensate fraction grows, leading to a transition of the density profile from a sandwich-type to a miscible configuration, as illustrated in the inset. The inset shows the density profiles at three representative temperatures, $T = 0\,T_c$, $0.4\,T_c$, and $0.8\,T_c$. In the main plot, the temperature is scaled by the critical temperature of the non-dipolar species}
    \label{mode_temp}
\end{figure}

 We now discuss the effects of thermal fluctuations present at finite temperatures on the quasiparticle excitation spectra. We consider  
\(N = 100\) atoms per unit area in each component and a fixed dipolar interaction strength of \(\varepsilon_{dd} = -0.35\), corresponding to attractive dipole-dipole interactions.
For finite-temperature calculations, the temperature is scaled by the critical temperature of the Rb component which is $T_c = 390\,\mathrm{nK}$. It is obtained by numerical self-consistent solution of GPEs and is the temperature at which the condensate fraction of the Rb component becomes nearly zero. At  zero temperature, the system resides in an immiscible regime with a sandwich configuration of the ground state. With an increase in thermal fluctuations, thermal atomic depletion becomes prominent, transforming the system toward a miscible phase with significant density overlap between the two components. It is worth noting that the thermal fluctuations-driven miscibility of binary non-dipolar mixtures has already been reported for continuum and lattice systems~\cite{roy_15,suthar_17,lingua_17,singh_23}. Moreover, the quantum fluctuations also play a crucial role in the miscibility of the mixture with impurity~\cite{bisset_21}. The mixing of dipolar and non-dipolar components with increasing temperature is evident from the ground-state density profiles shown in the inset of Fig.~\ref{mode_temp}, corresponding to $T = 0\,T_c$, $0.4\,T_c$, and $0.8\,T_c$.
At zero temperature, the attractive $\varepsilon_{dd}$ regime possesses three zero-energy modes corresponding to three distinct condensates of the sandwich immiscible configuration. We further find that the ground state remains robust with the self-consistent inclusion of quantum fluctuation present at zero temperature. As temperature is increased, the density dip in the non-dipolar component gets filled, and the peak of the dipolar component (at the center) lowers. At finite temperatures, the overall decrease in density is due to the presence of finite thermal densities (not shown here). As the system progresses towards miscibility with $T$, the change in the density profile of the ground-state condensate is attributed to the corresponding change in the low-lying spectra. In particular, we note that the softened (additional) zero-energy mode gains finite energy at the temperature of miscibility. This low-lying quasiparticle mode evolution with temperature is presented in Fig.~\ref{mode_temp}, and signals the loss of perfect phase separation and the onset of miscibility. 

The magnitude of the dipole mode amplitude increases as the two-component system approaches spatial mixing. Meanwhile, the in-phase dipole mode, associated with the center-of-mass motion of one species relative to the other, remains insensitive to the thermal effects. This is expected since such Kohn-type modes are predominantly governed by the external trapping potential and are less affected by internal interactions or thermal depletion~\cite{kohn_61,reidl_01}. Furthermore, the energy of the quadrupole mode, involving the shape oscillations and sensitive to the internal structure of the density distribution, changes non-monotonically with temperature. Initially, the mode softens with increasing temperature, and then at higher temperatures, the mode energy increases again, a behavior linked to enhanced thermal pressure and significant density overlap. Thus, the phase segregation of the dipolar and non-dipolar components of the mixtures is suppressed at finite temperatures. It is worth noting that the total number of atoms in each component, including both condensate and non-condensate, is conserved. Therefore, the observed modifications in the ground-state properties and quasiparticle spectra arise solely from finite-temperature effects and not caused by the change in the total number of atoms at higher temperatures. %We have finally explicitly checked that the change in the ground-state and quasiparticle spectra is solely due to non-zero temperatures and not due to lower particle numbers at finite $T$.
%%%%%%%%%%%%%%%%%%%%%%%%%%%%%%%%%%%%%%%%%%%%%%%%%%%%%%%%%%%%%%%%%%%%%%%%%%%%%%%%%%%%%%%%%%%%%%%%%%%%%%%%%%%%%%%%%%%%%%%%%%%%\subsection: correlation function%%%%%%%%%%%%%%%%%%%%%%%%%%%%%%%%%%%%%%%%%%%%%%%%%%%%%%%%%%%%%%%%%%%%%%%%%%%%%%%%%%%%%%%%%%%%%%%%%%%%%%%%%%%%%%%%%%%%%%%%%%%%%%%%%%%%%%%%%%%%%%%%%%%%%%%%%%%%%%%%%%%%
\subsection{First-Order Spatial Correlation Function}
We now examine the first-order spatial correlation function defined in terms of the off-diagonal condensate and noncondensate densities. The normalized correlation function can be expressed as
\begin{equation}
g^{(1)}_k(z, z') = \frac{n_{ck}(z,z') + \tilde{n}_k(z, z')}{\sqrt{n_k(z)\, n_k(z')}},
\label{eq:g1}
\end{equation}
where $n_{ck}(z,z') = \psi_k^*(z)\psi_k(z')$ is the off-diagonal condensate density, and $\tilde{n}_k(z,z') = \sum_j \{\big[ u_{kj}^*(z)\,u_{kj}(z') + v_{kj}^*(z)\,v_{kj}(z') \big] N_0(E_j) + v_{kj}^*(z)\,v_{kj}(z')\}$ is the two-point off-diagonal non-condensate densities of the $k$th component. At zero temperature, the entire condensate cloud has complete coherence, and therefore \( g^{(1)}_k = 1 \) within the condensate region. In binary condensates, the transition from phase-separated to the miscible domain at \( T \neq 0 \) has a characteristic signature in the spatial structure of the correlation function. 
\begin{figure}[h]
    \includegraphics[width=\linewidth]{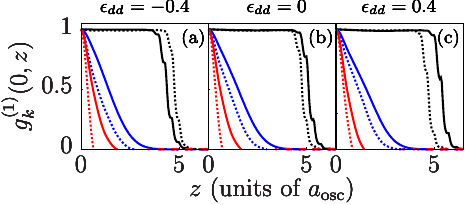}
    \caption{First-order correlation functions \( g^{(1)}_{\mathrm{Cr}}(0, z) \) (solid lines) and \( g^{(1)}_{\mathrm{Rb}}(0, z) \) (dashed lines) for a Cr-Rb mixture at equilibrium, shown for \( a_{12} = 7\,\mathrm{nm} \). The correlations are plotted for three different temperatures: \( T = 0 \) (black), \( 0.5\,T_c \) (blue), and \( 0.9\,T_c \) (red). The correlation functions are shown for three $\varepsilon_{dd}$ values mentioned at the top of the plots. Here, spatial coordinate $z$ is in units of oscillator length.} 
    \label{corr_fn}
\end{figure}
The first-order spatial correlation function \( g^{(1)}_{\mathrm{k}}(0, z) \) with temperature for various dipole-dipole interaction strengths is illustrated in Fig.~\ref{corr_fn}. At zero temperature (\(T = 0\,T_c\); black lines), the system exhibits near-perfect coherence across the bulk spatial extent of the condensates, with \( g^{(1)}_{\mathrm{k}}(0, z) \approx 1 \). The correlation drops at the periphery of the condensates. For attractive dipolar interactions (\( \varepsilon_{dd} < 0 \)), Cr condensate displays a slower decay of coherence compared to the non-dipolar component, i.e., \( g^{(1)}_{\mathrm{Rb}}(0,z) \) falls to zero over a longer distance than \( g^{(1)}_{\mathrm{Cr}}(0,z) \). This is attributed to the flanked Rb species in the periphery, and the Cr condensate occupies the center of the trap [cf. inset (a) of Fig.~\ref{mode_temp}]. As the temperature increases, the presence of non-condensate atoms tends to decrease the phase coherence present in the system. This results in a faster decay of \( g^{(1)}_{\mathrm{Cr}}(0, z) \) and \( g^{(1)}_{\mathrm{Rb}}(0, z) \) with distance. As shown in Fig.~\ref{corr_fn}, the black curves correspond to \( T = 0\,T_c \), where the coherence is nearly uniform across the condensate. As the temperature increases to \( 0.5\,T_c \) (blue curves) and further to \( 0.9\,T_c \) (red curves), the rate of decay increases significantly, indicating a reduction in long-range order due to thermal depletion and the growing contribution of non-condensate fractions. The correlation function is further fitted as $g^{(1)} \approx  \left[\alpha+e^{(\beta z - \gamma)}\right]^{-1}$ to capture the spatial characteristic of coherence. At zero temperature, the correlation acquires a steady value in the bulk (short distance from the center) with a decay at larger distances arising due to phase fluctuations. The parameter $\alpha$ signals the coherence amplitude in the bulk, $\beta$ controls the decay rate of correlation and is related to the inverse coherence length, and $\gamma$ determines the crossover length scale between the coherent and decaying regimes. With increasing temperature, the parameters $\alpha$ and $\gamma$ decrease, reflecting the reduction in coherence amplitude, while \(\beta\) increases, indicating a more rapid spatial decay of correlations. At finite $T$, the rapid decrease in $g^{(1)}(z)$ is attributed to the presence of thermal fluctuations at finite temperatures. 
%representing a nearly constant coherence in the bulk and then gradually drops to zero. With temperature, the parameters \(\alpha\) and \(\gamma\) adjust to reflect reduced coherence amplitude, while \(\beta\) increases, indicating a more rapid spatial decay of correlations. At finite $T$, the rapid decrease in $g^{(1)}(z)$ confirms the lower phase-coherent condensate regime with an increase in $T$. 

As $\varepsilon_{dd}$ drives a transition to a mixed state, at $\varepsilon_{dd} = 0$, the ground state of the system is in the miscible phase, and the absence of attractive dipolar interaction, Cr condensate atoms are pushed out, which signals a larger regime of coherence as compared to the non-dipolar Rb condensate, as shown in Fig.~\ref{corr_fn}(b). At repulsive dipolar interaction, the same effect is significant, as apparent from the solid and dashed black lines in Fig.~\ref{corr_fn}(c). At finite temperatures, the exponential decay of correlations from the center of the trap is similar to the attractive and zero $\varepsilon$ cases, with the decay rate increasing with $T$.  

%%%%%%%%%%%%%%%%%%%%%%%%%%%%%%%%%%%%%%%%%%%%%%% Dispersion relations %%%%%%%%%%%%%%%%%%%%%%%%%%%%%%%%%%%%%%%%%%%%%%%%%%%%%%%%%%%%%%%%%%%%%%%%%%%%%%%%%%%%%%%%
\subsection{Dispersion relations}
The perturbation of a uniform condensate exhibits phononic behaviour. Previous works have shown the two branches of in-phase and out-of-phase mode excitations at higher and lower energies, respectively~\cite{ticknor_14,momme_20}. These are associated with the fluctuations of the total density and spin density. To further analyze the character of the Bogoliubov excitations, we examine the dispersion relations of a quantum mixture of trapped dipolar and non-dipolar condensates. To this end, we perform the Fourier transform to obtain the quasiparticle amplitudes in momentum space and compute the expectation value of the momentum of the quasiparticles~\cite{wilson_10,ticknor_14}. For the $j$th quasiparticle mode, the expectation of the momentum in the $k$-space is  
\begin{equation}
\mathcal{K}^{\mathrm{rms}}_{j} = \left(\frac{\displaystyle \sum_{k}\int \! d\mathbf{q}\; |\mathbf{q}|^{2}\left[\,|{u}_{k j}(\mathbf{q})|^{2}+|{v}_{k j}(\mathbf{q})|^{2}\,\right]
}{\displaystyle \sum_{k}\int \! d\mathbf{q}\; \left[\,|{u}_{k j}(\mathbf{q})|^{2}+|{v}_{k j}(\mathbf{q})|^{2}\,\right]}
\right)^{1/2}.
\end{equation}
\begin{figure}[h]
    \includegraphics[width=\linewidth]{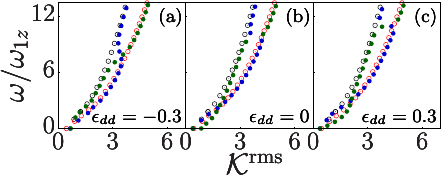}
    \caption{The discrete quasiparticle dispersion relation for the $^{52}\mathrm{Cr}$–$^{87}\mathrm{Rb}$ binary mixture at zero temperature. The filled circles represent the dispersion relation for binary mixture of dipolar and non-dipolar condensates at $a_{12} = 7~\mathrm{nm}$. For reference, the dispersion curves for a mixture of non-dipolar condensates with zero intercomponent interactions are shown with open (unfilled) circles. The curves are shown for $\varepsilon_{dd} = -0.3$, $0$, and $0.3$. The blue filled circles correspond to the in-phase quasiparticle excitations, whereas the green filled circles represent the out-of-phase excitations. Here, the black and red open circles correspond to first and second species, respectively.}
  \label{dispersion_reln}
\end{figure}
We show the zero-temperature dispersion relations for different strengths of the dipolar interactions in Fig.~\ref{dispersion_reln}. The low-energy excitations exhibit a phonon-like linear behavior at small momenta. In the absence of intercomponent interaction, the dispersion curves exhibit two distinct branches corresponding to each of the species. Here, we identify the upper and lower branches that correspond to the first and second species, shown by open circles in Fig.~\ref{dispersion_reln}(a,b,c). It is worth noting that the mode mixing occurs due to finite intercomponent interactions. Here, as the dipolar strength remains lower to the short-range contact interactions $\varepsilon_{dd}<1$, and following momentum-sector decoupling dipolar interaction is approximated to a contact-like form with zero in-plane momentum, we do not observe the roton-minimum in the dispersion curves~\cite{petter_19}. In Fig.~\ref{dispersion_reln}(a), the ground state density profile for $\varepsilon_{dd} = -0.3$ is of sandwich-type configuration; the lower branch (blue-filled circle) corresponds to the in-phase quasiparticle oscillations, while the upper branch (green-filled circle) represents out-of-phase excitations. As the overlap of the condensates is minimal, the mixing of the quasiparticle modes leads to crossing of the branches. We further find that as branches approach close proximity, the mode hybridization takes place, and at higher energies the curve shows a transition from excitation of in-phase character to out-of-phase character and vice versa. For $\varepsilon_{dd} = 0$, when the system begins to transition from the immiscible to the miscible regime, the in-phase mode excitations have higher energy than that of out-of-phase excitations, as evident from Fig.~\ref{dispersion_reln}(b). The mode hybridization and crossing of the branches occur at higher momenta. The Fig.~\ref{dispersion_reln}(c) represents the branch crossing point further shifts to higher momenta, which signals lower energies for low-lying out-of-phase quasiparticle excitations. It is important to note that the character of the low-lying modes changes with different $\varepsilon_{dd}$. In particular, for attractive dipolar interactions $\varepsilon_{dd}<0$, the in-phase modes have lower quasiparticle energies, while at zero or repulsive interaction $\varepsilon_{dd}\gtrsim 0$, the out-of-phase modes have lower quasiparticle energies. This is corresponding to the quantum phase transition from the immiscible to miscible ground state of the mixture. Thus, the energy hierarchy of in-phase and out-of-phase modes changes with $\varepsilon_{dd}$, which is consistent with the quasiparticle mode evolution shown in Fig.~\ref{mode_a12_7nm}. The temperature-dependent mode evolution corresponding to the mixing transition also exhibits similar dispersion relations.
%%%%%%%%%%%%%%%%%%%%%%%%%%%%%%%%%%%%%%%%%%%%%%%%%%%%%%%%%%%%%%%%%%%%%%%%%%%
%%%%      Conclusions                                          %%%%%%%%%%%
%%%%%%%%%%%%%%%%%%%%%%%%%%%%%%%%%%%%%%%%%%%%%%%%%%%%%%%%%%%%%%%%%%%%%%%%%%%
\section{CONCLUSIONS}\label{CONCLUSIONS}
 We have examined the collective modes of a binary mixture of dipolar and non-dipolar Bose-Einstein condensates at zero and finite temperatures. For a highly flattened infinite pancake configuration with uniform radial atomic density, the dipolar interaction reduces to a contact-like form, where the magnetic dipoles are polarized along the $z$-axis. We present intriguing
low-lying quasiparticle mode evolutions driven by the dipolar interaction and temperature. At zero temperature, the ground state is an immiscible phase in an attractive dipolar interaction $(\varepsilon < 0)$, which corresponds to three zero-energy modes in the excitation spectra, including an additional Goldstone mode of a phase-separated sandwich state. The tuning of the dipolar interaction from attractive to repulsive results in a hardening of the additional zero-energy mode, which is concomitant with the quantum phase transition from immiscible to miscible phases. Nevertheless, with increased repulsive interactions between components (resulting from a larger number of atoms), the spectra of low-lying modes display discontinuities that correspond to the exchange of species positions.  

The thermal fluctuations at finite temperatures also mix the components with attractive dipolar interactions. This is evident in the reappearance of the dipole mode at $T\approx 0.6~T_{c}$, with $T_{c}$ representing the critical temperature of condensation for the dipolar component. We further analyze the degree of phase coherence and the decay rate of the first-order correlation function. The steady value of correlation in the bulk indicates the presence of phase coherence at zero temperature, while it decays sharply at finite temperatures due to loss of phase coherence. Furthermore, the low-lying in-phase (out-of-phase) modes appear in the lower branch of the dispersion curve for the attractive (zero or repulsive) dipolar interactions. The mode hybridization or crossing of branches of the dispersion curve shifts to higher momenta as the dipolar interaction is tuned from attractive to repulsive. Our study emphasizes the significance of anisotropic dipolar interactions in the evolution of low-lying axial quasiparticle modes within the infinite pancake regime. This further opens avenues for exploring the collective excitations of the ferrofluid phase in higher-dimensional quantum systems.

\begin{acknowledgments}
H.K. acknowledges the financial support from University Grant Commission (UGC), New Delhi. K.S. acknowledges support from the Science and Engineering Research Board, Department of Science and Technology, Government of India through Project No. SRG/2023/001569.
\end{acknowledgments}
%%%%%%%%%%%%%%%%%%%%%%%%%%%%%%%%%%%%%%%%%%%%%%%%%%%%%%%%%%%%%%%%%%%%%%%%%%%

\bibliography{dip}{}
\bibliographystyle{apsrev4-1}
\end{document}